\documentclass[twocolumn,prl,showpacs,amsmath,amssymb]{revtex4}

\usepackage{graphicx}
\usepackage{color}
\usepackage{dcolumn}
\usepackage{bm}
\include{epsf}

\begin{document}

\title{Optical Measurement of the Phase-Breaking Length in Graphene}

\author{Ryan Beams$^{1}$, Luiz Gustavo Can\c{c}ado$^{2}$, and Lukas Novotny$^{1}$}
\homepage{www.nanooptics.org}
\affiliation{$^{1}$Institute of Optics, University of Rochester, Rochester, New York 14627, USA}
\affiliation{$^{2}$Departamento de F\'{i}sica, Universidade Federal de Minas Gerais, Belo Horizonte, MG 30123-970, Brazil}

\date{\today}

\begin{abstract}
In mesoscopic physics, interference effects play a central role on the transport properties of conduction electrons, giving rise to exotic phenomena such as weak localization, Aharonov-Bohm effect, and universal conduction fluctuations. Mesoscopic objects have a size on the order of the {\em phase-breaking length} $L_{\phi}$, the length conduction electrons  travel while keeping phase coherence. In this letter, we use vibrational spectroscopy in combination with a novel optical defocusing method to measure $L_{\phi}$ of photo-excited electrons in graphene which undergo inelastic scattering by optical phonons.  We extract $L_{\phi}$ from the spatial confinement of the defect-induced Raman D band near the edges of graphene. Temperature dependent measurements in the range of 1.55\,K to 300\,K  yield  $L_{\phi} \propto 1/\sqrt{T}$, in agreement with previous magneto-transport measurements.
\end{abstract}

\maketitle

Over the past decades, silicon based materials spurred the development of ever smaller and faster transistor devices and the production of ever larger information storage. However, as the material dimensions become comparable to the length scale of electronic wavefunctions it is necessary to explore alternative materials with enhanced performance. Low-dimensional carbon materials, such as carbon nanotubes and graphene,  form a particularly promising alternative to silicon based materials~\cite{avouris2007,castro2010}.
Monolayer graphene exhibits unique electronic transport properties, such as high electron mobility at room temperature ($\geq$\,10$^{4}$\,cm$^{2}$/Vs) and tunable carrier densities as high as 10$^{13}$\,cm$^{-2}$~\cite{novoselov2007}. These properties have led to extensive research and motivated the development  of proof-of-concept devices, such as single electron transistors (SETs), field-effect transistors (FETs), and even graphene-based memory~\cite{ensslin2008,meric2008,tour2008}. Although single-crystal graphene is metallic (unable to confine electrons electrostatically), it can become semiconducting when produced in a narrow ($\leq$\,10\,nm) ribbon-like geometry, a manifestation
of electronic quantum confinement~\cite{avouris2007}.\\[-3ex]

Conduction electrons confined to nanometer length scales cease to show a purely diffusing behavior as predicted by the Boltzmann transport theory. In this regime, the dephasing length of conduction electrons becomes larger than the device itself, and interference effects give rise to weak localization, Aharonov-Bohm effect, and universal conduction fluctuations~\cite{morozov2006,wu2007,tikho2009,ludeberg2009}.
These interference effects extend over the length $L_{\phi}$, the electron {\em phase-breaking length}~\cite{taylor2002,been1991}. In this letter, we use Raman scattering to experimentally determine the  {\em phase-breaking length} of photo-excited electrons near the edges of graphene, where electrons undergo inelastic scattering with optical phonons. The measurements were performed using a novel optical defocusing  method, which makes it possible to measure the localization of the disorder-induced Raman D band~\cite{tuinstra1969} with a resolution of a few nanometers.\\[-2ex]

The relationship between the spatial confinement of the D band and the {\em phase-breaking length} has been debated recently~\cite{cancado2008B,casiraghi2009}.  In one interpretation, D band localization is predicted to depend on the virtual electron-hole lifetime and $L_{\phi}$ on the electron-phonon scattering rate~\cite{casiraghi2009}. According to this interpretation, there is no direct relationship between D band localization and $L_{\phi}$. In contrast, another interpretation predicts that D band localization is a direct consequence of the phase-breaking length and that $L_{\phi}$ can be measured by determining the spatial confinement of the D band~\cite{cancado2008B}. To resolve this controversy, we have performed temperature-dependent measurements of D band scattering near the edges of graphene. As the temperature is decreased from 300\,K to 1.55\,K, the phonon population decreases and  $L_{\phi}$ increases~\cite{been1991,morozov2006,wu2007,tikho2009}. Our measurements reveal that $L_{\phi}$ varies as $1/\sqrt{T}$, in agreement with magneto-transport measurements~\cite{morozov2006}.\\[-2ex]

\begin{figure}[!t]
\includegraphics[width=21em]{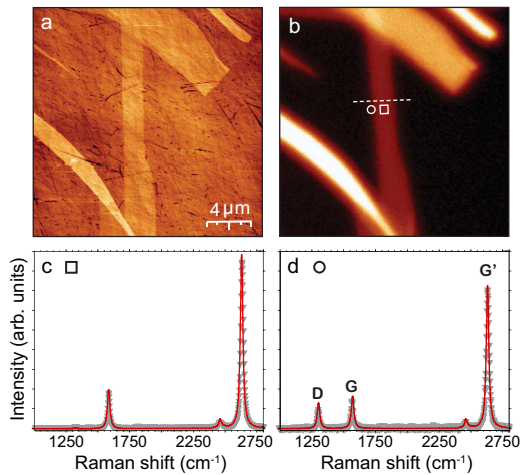}
\caption{(a) Topographic image of exfoliated graphene deposited on a fused silica substrate. (b) Confocal Raman image showing the intensity of the G band over the same region. The dashed line indicates the direction along which Raman scattering spectra were recorded. (c,d) Raman spectra evaluated at the locations indicated by the square and circle in (b). The D band shows up only at the graphene edge.  Spectrum (d) was scaled by a factor of 3 relative to spectrum (c). \label{confocal}}
\end{figure}
Figure~\ref{confocal}(a) shows a topographic image recorded by atomic force microscopy (AFM) of a set of graphene pieces deposited on a fused silica surface. A corresponding confocal Raman image of the same region is depicted in Figure~\ref{confocal}(b). The contrast (color scale) in the Raman image renders the intensity of the bond-stretching G band. The G band is a first-order scattering process that originates from the double-degenerate vibrational mode $\Gamma_{6}^{+}$ (E$_{2g}$) that occurs at the crossing of the longitudinal optical (LO) and transverse optical (TO) phonon branches at the $\Gamma$ point in the 1$^{st}$ Brillouin zone of graphene~\cite{tuinstra1969}. Figures~\ref{confocal}(c,d) show full Raman scattering spectra acquired at two different locations indicated by the white square and circle in Fig.~\ref{confocal}(b).  Spectrum (c) was acquired from the interior of the graphene piece ($\approx 1\,\mu$m from the edge), whereas spectrum (d) was recorded with the incident laser focus centered at the left edge. Spectrum (c) shows two major Raman features: the G band occurring at $\simeq$\,1580\,cm$^{-1}$, and the G$^{\,\prime}$ band centered at $\simeq$\,2700\,cm$^{-1}$. The G$^{\,\prime}$ band originates from two-phonon Raman processes involving TO phonons~\cite{ferrari2001} whose wavevectors extend next to the corner of the 1$^{st}$ Brillouin zone ($K$ and $K^{\prime}$ points). The single-Lorentzian shape of the G$^{\prime}$ band indicates that the measured sample region is indeed a single-layer graphene piece~\cite{ferrari2006,cancado2008}.\\[-2ex]

Both G and G$^{\,\prime}$ bands are also detected in the Raman spectrum of Figure~\ref{confocal}(d) (obtained from the edge). In addition to the peaks identified in the previous spectrum, we now also observe the defect-induced D band centered at $\simeq$\,1350\,cm$^{-1}$~\cite{tuinstra1969}. Due to momentum conservation, the TO phonons giving rise to this band only become Raman active if the electrons or holes involved in the scattering process undergo elastic scattering by a lattice defect, which in this case is provided by the graphene edge~\cite{cancado2004,cancado2008B,casiraghi2009}. Therefore, the closer the position of a photo-excited electron to the graphene edge, the higher is its probability to be involved in D band scattering. Notice that, although the G$^{\,\prime}$ band is the overtone of the D band, it does not require the presence of defects to become Raman active, since momentum conservation is guaranteed in two-phonon Raman processes. The Raman spectra shown in Figures~\ref{confocal}(c,d) were recorded with the polarization vector of the excitation laser beam oriented parallel to the graphene edge, since polarizations perpendicular to the edge do not generate D band scattering~\cite{cancado2004,casiraghi2009}.\\[-2ex]
			
\begin{figure}[!b]
\includegraphics[height=19em]{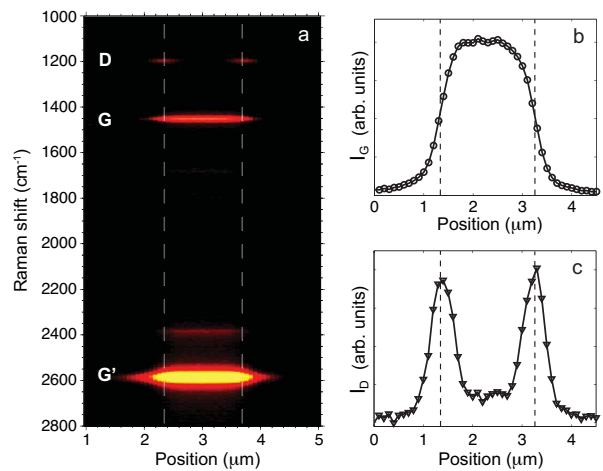}
\caption{(a) Hyperspectral line scan recorded along the dashed line in Figure~\ref{confocal}(b). The vertical lines indicate the position of the edges of the graphene piece. The D band is localized at the graphene edges.  (b,c) Intensity profiles of the G and D bands obtained from the hyperspectral data shown in (a). \label{hyper}}
\end{figure}

In order to analyze the spatial dependence of the D, G, and G$^{\,\prime}$ intensities we have recorded Raman scattering spectra for different positions of the laser focus relative to the graphene edge. The incident beam was moved in steps of 100\,nm along a 5\,$\mu$m long line perpendicular to the edges of a narrow graphene piece [dashed line in Figure~\ref{confocal}(b)]. Figure~\ref{hyper}(a) shows a hyperspectral Raman map recorded along the scan line. The horizontal axis refer to the spatial position of the incident laser (in $\mu$m units) and the vertical axis renders the Raman shift. As the graphene piece is stepped through the laser focus, the intensities of the G and G$^{\,\prime}$ bands gradually transit from a minimum value (dark counts) to a maximum value. On the other hand, the D band intensity achieves a maximum value when the graphene edge is in the laser focus. Figures~\ref{hyper}(b)~and~\ref{hyper}(c) show the intensity profile of the G and D bands, respectively, obtained from the hyperspectral data shown in panel (a).\\[-2ex]

Because the intensity profile of the incident laser beam corresponds to a finite point-spread function (PSF), the intensity profiles shown in Figs.~\ref{hyper}(a-c) originate from the spatial convolution of the Raman susceptibilities with the strength of the incident electric field. Initial attempts to determine the spatial confinement $\ell_{\rm D}$ of the D band have been made by making use of prior information about the PSF of a strongly focused beam~\cite{cancado2008B}. This prior knowledge was used to deconvolve the D band confinement from the recorded scan profile across a graphene edge. However, this method was only able to provide an upper limit of about 20\,nm~\cite{cancado2008B} and is not accurate enough to quantitatively determine the D band confinement. To overcome previous limitations we here introduce a defocusing technique  to extract $\ell_{\rm D}$ from the ratio of G and D band intensities.\\[-2ex]

\begin{figure}[!b]
\includegraphics[height=15.5em]{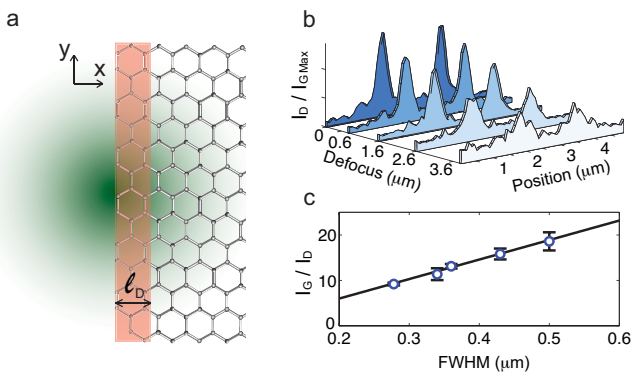}
\caption{(a)  Sketch of the measurement technique. The laser focus is scanned across the graphene edge along the $x$ direction. The light red rectangle indicates the region in which the D band signal is generated.  (b) Line scan profile for the ratio $I_{\rm D}/I_{\rm G_{Max}}$ obtained along the dashed line shown in Figure~\ref{confocal}({\bf a}) for different amounts of defocus.  (c) Plot of the experimental data of the ratio $I_{\rm G}/I_{\rm D}$ as a function of $\Delta$. The solid line is a linear fit giving $I_{\rm G}/I_{\rm D}\,=\,$42\,${\mu}$m$^{-1}$\,$\Delta$. \label{defocus}}
\end{figure}
When the laser focus is centered at the edge of the graphene piece, the G band intensity ($I_{\rm G}$) is proportional to the area of the laser focus, whereas the D band intensity is zero in absence of defects. D band scattering can be observed once the laser focus is placed near a graphene edge, which serves as a one-dimensional model defect. As depicted in Figure~\ref{defocus}(a), the D band intensity ($I_{\rm D}$) is proportional to the area defined by the spatial D band confinement $\ell_{\rm D}$ (measured from the edge) and the length of the graphene edge exposed to the incident laser beam. Hence, the intensity ratio $I_{\rm G}/I_{\rm D}$ turns out to be proportional
to the full-width-at-half-maximum (FWHM) of the point-spread function. For a Gaussian focus of width $\sigma$, this result can be formally derived as
\begin{equation}\label{int01}
I_{\rm G} \propto\int_{0}^{\infty }\int_{-\infty }^{\infty}\chi_{\rm G}^{2}\;e^{-\frac{(x^2 + y^2)}{2\sigma^2}}\,dy\:\!dx\,=\,\pi\chi_{\rm G}\sigma^{2}\,,
\end{equation}
\begin{equation}\label{int02}
I_{\rm D}\propto\int_{0}^{\ell_{\rm D}}\int_{-\infty }^{\infty}\chi_{\rm D}^{2}\;e^{-\frac{(x^2 + y^2)}{2\sigma^2}}\,dy\:\!dx\,\simeq\,\sqrt{2\pi}\,\ell_{\rm D}\,\chi_{\rm D}\sigma\,,
\end{equation}
where the lateral position of the edge is fixed at $x$\,=\,0  (see Fig.~\ref{defocus}a for reference). $\chi_{\rm G}$ and $\chi_{\rm D}$ are the Raman susceptibilities of the G and D bands, respectively. Notice that we have considered $\ell_{\rm D}\ll\sigma$ in Eq.~(\ref{int02}). The ratio $I_{\rm G}/I_{\rm D}$ obtained from Eqs.~(\ref{int01})~and~(\ref{int02}) for the focus centered at the edge gives
\begin{equation}\label{ratio01}
\frac{I_{\rm G}}{I_{\rm D}} = \frac{1}{4}\sqrt{\frac{\pi}{ln(2)}}\frac{\chi_{\rm G}^2}{\chi_{\rm D}^2}\frac{\Delta}{\ell_{\rm D}}\,,
\end{equation}
where we have used the relation $\sigma\,=\,\Delta/[2\sqrt{2\,\ln(2)}]$, with $\Delta$ being the FWHM. $\chi_{\rm G}^2/\chi_{\rm D}^2$ is the ratio between the strengths of the G and D bands in graphene, and has been empirically determined in Ref.~\cite{lucchese2010} to have a value of $\sim$\,0.24.\\[-2ex]

According to equation~(\ref{ratio01}), the plot of the ratio $I_{\rm G}/I_{\rm D}$ versus $\Delta$ gives a line whose slope is proportional to $\ell_{\rm D}$. Therefore, the width $\ell_{\rm D}$ can be experimentally determined by performing several line scans with different amounts of defocus, which means different $\Delta$ values. Since $\ell_{\rm D}\,\ll\,\sigma$, as shown in Ref.~\cite{cancado2008B}, the spatial profile of the D band is a good approximation to the line response function of the focus and can be used as a direct measurement of $\Delta$.  The ratio $I_{\rm D}/I_{\rm G_{Max}}$ for different amount of defocus is shown in Fig~\ref{defocus}(b), where $I_{\rm G_{Max}}$ is the $I_{\rm G}$ value taken from the center of the sample. Figure~\ref{defocus}(c) shows the plot $I_{\rm G}/I_{\rm D}$ versus $\Delta$ for the experimental data obtained using this technique at room temperature. The linear fit (solid line) gives $I_{\rm G}/I_{\rm D}\,=\,$42\,${\mu}$m$^{-1}$\,$\Delta$, which according to equation~(\ref{ratio01}) leads to $\ell_{\rm D}\,\approx\,$3\,nm. This is in excellent agreement with recent work by Lucchese {\em et al.} where the authors have measured a D band relaxation length of $\approx$\,3\,nm, for point defects in ion bombarded graphene samples~\cite{lucchese2010}.\\[-2ex]

Let us now discuss the physical origin of the D band confinement $\ell_{\rm D}$. A recent work~\cite{basko2009} have explained that for {\em any} one-phonon Raman scattering process (such as D band scattering) the relaxation length is associated with the virtual electron-hole pair lifetime.  The authors argued that since the energy uncertainty in such a process is determined by the  phonon energy $\omega_{ph}$, the minimum value for the electron-hole pair lifetime obtained from the uncertainty principle is 1/$\omega_{ph}$. By using the relation $v/\omega_{ph}$, where $v$ is the group velocity of the photo-excited electrons, the authors have found $\ell_{\rm D}\,\approx\,$4\,nm. This value is in agreement with the result obtained here, and also in Ref.~\cite{lucchese2010}. However, this picture only gives an estimate for the minimum scattering length for the D band and does not explain the underlying physical mechanisms. \\[-2ex]

\begin{figure}[!b]
\begin{center}
\includegraphics[height=21em]{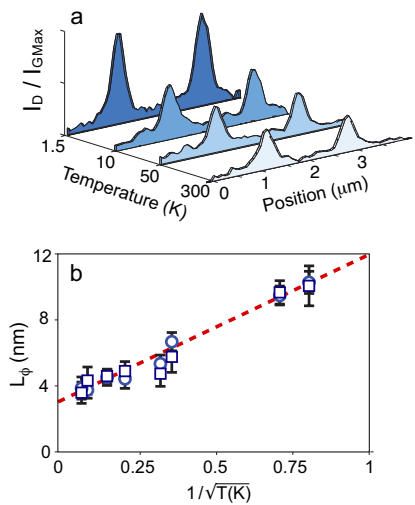}
\caption{(a) Scan line profile for the ratio $I_{\rm D}/I_{\rm G_{Max}}$ along the dashed line in Figure~\ref{confocal}(b) evaluated for different temperatures. (b) Plot of $L_{\phi}$ as a function of $1/\sqrt{T}$ for the left edge (blue circles) and the right edge (red squares) of the graphene sample. The solid line shows a linear fit giving $L_{\phi}{\rm(nm)}\,=\,3\,+\,9/\sqrt{T}$. \label{temp}}
\end{center}
\end{figure}

The {\em phase-breaking length} $L_{\phi}$ of a conduction electron is defined as the average distance traveled before undergoing inelastic scattering~\cite{taylor2002,been1991}. Since electrons involved in D band Raman scattering undergo a single inelastic scattering event with a lattice phonon, $L_{\phi}$ must correspond to the average distance traveled by such electrons during the time interval in which the D band scattering process takes place, that is $\ell_{\rm D}$\,=\,$L_{\phi}$. By cooling the graphene sample we are able to reduce the phonon population and to decrease the probability of electrons to inelastically scatter by phonons. Thus, $L_{\phi}$ should increase as the temperature $T$ decreases. To verify this hypothesis, we have collected Raman scattering spectra for temperatures ranging from 1.55\,K to 300\,K.  In Figure~\ref{temp}(a) we plot the ratio $I_{\rm D}/I_{\rm G_{Max}}$ for different values of temperature along the scan line indicated in Figure~\ref{confocal}(b).  The experimental data show that the ratio increases as the temperature decreases (c.f. Figure~\ref{temp}a), indicating that the D band scattering length is indeed increasing. This result provides clear evidence for D band delocalization at low temperatures. \\[-2ex]

To quantify our experimental data and to relate the measurements to $L_{\phi}$ we have recorded the temperature dependence of $\ell_{\rm D}$ by repeating the defocusing experiment in Figure~\ref{defocus}(c) for different temperatures. The results are shown in Figure~\ref{temp}(b), which depicts $L_{\phi}$ as a function of inverse sample temperature for both edges. The solid line is a linear fit showing that $L_{\phi}\,\propto\,1/\sqrt{T}$. Since $L_{\phi}$ is known to be inversely proportional to $\sqrt{T}$~\cite{been1991,morozov2006}
our result provides strong evidence for the hypothesis that the spatial confinement of the D band indeed reflects the {\em phase-breaking length}. \\[-2ex]

The linear fit in Figure~\ref{defocus}(b) yields $L_{\phi}{\rm(nm)}\,=\,3\,+\,9/\sqrt{T}$, with a limiting value of $L_{\phi}$= 3\,nm for high temperatures ($T\,\geq$\,300\,K). This result corroborates with our assumption that the value estimated for $\ell_{\rm D}$ in Ref.\rm{\cite{casiraghi2009}} corresponds to the low limit for  $L_{\phi}$ determined by the uncertainty principle. This limit is accurate for high phonon population regimes.\\[-2ex]

In summary, we have introduced a novel optical defocusing method for the measurement of the spatial confinement and temperature dependence of localized states, such as D band Raman scattering. Our measurements yield a D band confinement of $\ell_{\rm D}\,=\,3\,$nm at T=300\ K and a temperature dependence of $1/\sqrt{T}$, which identifies $\ell_{\rm D}$ as the {\em phase-breaking length} $L_{\phi}$ of photo-exited electrons or holes that undergo inelastic scattering by optical phonons. The technique developed here has clear advantages over standard transport measurements because contributions due to different scattering processes (electron-electron, electron-phonon, and electron-impurity) are not intermixed. The work demonstrates a strong connection between electronic and optical properties of graphene and shows that Raman spectroscopy provides an alternative tool for studying electric transport in mesoscopic structures.\\[-2ex]

\begin{acknowledgments}
The authors would like to thank Christiane H{\"o}ppener, Bradley Deutsch, Palash Bharadwaj, and Ado Jorio for valuable discussions and input. This work was supported by the Department of Energy (grant DE-FG02-05ER46207).
\end{acknowledgments}


\begin{thebibliography}{30}
\expandafter\ifx\csname natexlab\endcsname\relax\def\natexlab#1{#1}\fi
\expandafter\ifx\csname bibnamefont\endcsname\relax
  \def\bibnamefont#1{#1}\fi
\expandafter\ifx\csname bibfnamefont\endcsname\relax
  \def\bibfnamefont#1{#1}\fi
\expandafter\ifx\csname citenamefont\endcsname\relax
  \def\citenamefont#1{#1}\fi
\expandafter\ifx\csname url\endcsname\relax
  \def\url#1{\texttt{#1}}\fi
\expandafter\ifx\csname urlprefix\endcsname\relax\def\urlprefix{URL }\fi
\providecommand{\bibinfo}[2]{#2}
\providecommand{\eprint}[2][]{\url{#2}}

\bibitem[{\citenamefont{Avouris et~al.}(2007)\citenamefont{Avouris, Chen, and
  Perebeinos}}]{avouris2007}
\bibinfo{author}{\bibfnamefont{P.}~\bibnamefont{Avouris}},
  \bibinfo{author}{\bibfnamefont{Z.}~\bibnamefont{Chen}}, \bibnamefont{and}
  \bibinfo{author}{\bibfnamefont{V.}~\bibnamefont{Perebeinos}},
  \bibinfo{journal}{Nature Nanotech.} \textbf{\bibinfo{volume}{2}},
  \bibinfo{pages}{605} (\bibinfo{year}{2007}).

\bibitem[{\citenamefont{Castro-Neto}(2010)}]{castro2010}
\bibinfo{author}{\bibfnamefont{A.~H.} \bibnamefont{Castro-Neto}},
  \bibinfo{journal}{Materials Today} \textbf{\bibinfo{volume}{13}},
  \bibinfo{pages}{12} (\bibinfo{year}{2010}).

\bibitem[{\citenamefont{Geim and Novoselov}(2007)}]{novoselov2007}
\bibinfo{author}{\bibfnamefont{A.~K.} \bibnamefont{Geim}} \bibnamefont{and}
  \bibinfo{author}{\bibfnamefont{K.~S.} \bibnamefont{Novoselov}},
  \bibinfo{journal}{Nature Phys.} \textbf{\bibinfo{volume}{6}},
  \bibinfo{pages}{183} (\bibinfo{year}{2007}).

\bibitem[{\citenamefont{Novoselov et~al.}(2004)\citenamefont{Novoselov, Geim,
  Morozov, Jiang, Zhang, Dubono, Grigorieva, and Firsov}}]{novoselov2004}
\bibinfo{author}{\bibfnamefont{K.~S.} \bibnamefont{Novoselov {\it et al.}}},
\bibinfo{journal}{Science}
  \textbf{\bibinfo{volume}{306}}, \bibinfo{pages}{666} (\bibinfo{year}{2004}).

\bibitem[{\citenamefont{Berger et~al.}(2004)\citenamefont{Berger, Zhimin, Li,
  Li, Ogbazghi, Feng, Dai, Marchenkov, Conrad, First et~al.}}]{berger2004}
\bibinfo{author}{\bibfnamefont{C.}~\bibnamefont{Berger {\it et al.}}},
\bibinfo{journal}{J. Phys. Chem. B}
  \textbf{\bibinfo{volume}{108}}, \bibinfo{pages}{19912}
  (\bibinfo{year}{2004}).

\bibitem{ensslin2008}
C. Stampfer {\it et al.}, 
Nano Lett. {\bf 8}, 2378 (2008).

\bibitem[{\citenamefont{Meric et~al.}(2008)\citenamefont{Meric, Han, Youg,
  Ozyilmaz, Kim, and Shepard}}]{meric2008}
\bibinfo{author}{\bibfnamefont{I.}~\bibnamefont{Meric} {\it et al.}},
  \bibinfo{journal}{Nature Nanotech.} \textbf{\bibinfo{volume}{3}},
  \bibinfo{pages}{654} (\bibinfo{year}{2008}).

\bibitem[{\citenamefont{Li et~al.}(2008{\natexlab{a}})\citenamefont{Li,
  Sinitshii, and Tour}}]{tour2008}
\bibinfo{author}{\bibfnamefont{Y.}~\bibnamefont{Li}},
  \bibinfo{author}{\bibfnamefont{A.}~\bibnamefont{Sinitshii}},
  \bibnamefont{and} \bibinfo{author}{\bibfnamefont{J.~M.} \bibnamefont{Tour}},
  \bibinfo{journal}{Nature Mat.} \textbf{\bibinfo{volume}{7}},
  \bibinfo{pages}{966} (\bibinfo{year}{2008}{\natexlab{a}}).

\bibitem[{\citenamefont{Levendorf et~al.}(2009)\citenamefont{Levendorf,
  Ruiz-Vargas, Garg, and Park}}]{levendorf2009}
\bibinfo{author}{\bibfnamefont{M.~P.} \bibnamefont{Levendorf}},
  \bibinfo{author}{\bibfnamefont{C.~S.} \bibnamefont{Ruiz-Vargas}},
  \bibinfo{author}{\bibfnamefont{S.}~\bibnamefont{Garg}}, \bibnamefont{and}
  \bibinfo{author}{\bibfnamefont{J.}~\bibnamefont{Park}},
  \bibinfo{journal}{Nano Lett.} \textbf{\bibinfo{volume}{9}},
  \bibinfo{pages}{4479} (\bibinfo{year}{2009}).

\bibitem[{\citenamefont{Li et~al.}(2008{\natexlab{b}})\citenamefont{Li, Wang,
  Zhang, Lee, and Dai}}]{dai2008}
\bibinfo{author}{\bibfnamefont{X.}~\bibnamefont{Li}},
  \bibinfo{author}{\bibfnamefont{X.}~\bibnamefont{Wang}},
  \bibinfo{author}{\bibfnamefont{L.}~\bibnamefont{Zhang}},
  \bibinfo{author}{\bibfnamefont{S.}~\bibnamefont{Lee}}, \bibnamefont{and}
  \bibinfo{author}{\bibfnamefont{H.}~\bibnamefont{Dai}},
  \bibinfo{journal}{Science} \textbf{\bibinfo{volume}{319}},
  \bibinfo{pages}{1229} (\bibinfo{year}{2008}{\natexlab{b}}).

\bibitem[{\citenamefont{Dubois et~al.}(2009)\citenamefont{Dubois, Zanolli,
  Declerck, and Charlier.}}]{charlier2009}
\bibinfo{author}{\bibfnamefont{S.~M.~M.} \bibnamefont{Dubois}},
  \bibinfo{author}{\bibfnamefont{Z.}~\bibnamefont{Zanolli}},
  \bibinfo{author}{\bibfnamefont{X.}~\bibnamefont{Declerck}}, \bibnamefont{and}
  \bibinfo{author}{\bibfnamefont{J.~C.} \bibnamefont{Charlier.}},
  \bibinfo{journal}{Eur. Phys. J. B} \textbf{\bibinfo{volume}{72}},
  \bibinfo{pages}{1} (\bibinfo{year}{2009}).

\bibitem[{\citenamefont{Morozov et~al.}(2006)\citenamefont{Morozov, Novoselov,
  Schedin, Ponomarenko, Jiang, and Geim}}]{morozov2006}
\bibinfo{author}{\bibfnamefont{S.~V.} \bibnamefont{Morozov} {\it et al.}},
  \bibinfo{journal}{Phys. Rev. Lett.} \textbf{\bibinfo{volume}{97}},
  \bibinfo{pages}{016801} (\bibinfo{year}{2006}).

\bibitem[{\citenamefont{Wu et~al.}(2007)\citenamefont{Wu, Li, Song, Berger, and
  de~Heer}}]{wu2007}
\bibinfo{author}{\bibfnamefont{X.}~\bibnamefont{Wu}},
  \bibinfo{author}{\bibfnamefont{X.}~\bibnamefont{Li}},
  \bibinfo{author}{\bibfnamefont{Z.}~\bibnamefont{Song}},
  \bibinfo{author}{\bibfnamefont{C.}~\bibnamefont{Berger}}, \bibnamefont{and}
  \bibinfo{author}{\bibfnamefont{W.~A.} \bibnamefont{de~Heer}},
  \bibinfo{journal}{Phys. Rev. Lett.} \textbf{\bibinfo{volume}{98}},
  \bibinfo{pages}{136801} (\bibinfo{year}{2007}).

\bibitem[{\citenamefont{Tikhonenko et~al.}(2009)\citenamefont{Tikhonenko,
  Kozikov, Savchenko, and Gorbachev}}]{tikho2009}
\bibinfo{author}{\bibfnamefont{F.~V.} \bibnamefont{Tikhonenko}},
  \bibinfo{author}{\bibfnamefont{A.~A.} \bibnamefont{Kozikov}},
  \bibinfo{author}{\bibfnamefont{A.~K.} \bibnamefont{Savchenko}},
  \bibnamefont{and} \bibinfo{author}{\bibfnamefont{R.~V.}
  \bibnamefont{Gorbachev}}, \bibinfo{journal}{Phys. Rev. Lett.}
  \textbf{\bibinfo{volume}{103}}, \bibinfo{pages}{226801}
  (\bibinfo{year}{2009}).



\bibitem[{\citenamefont{Lundeberg and Folk}(2009)}]{ludeberg2009}
\bibinfo{author}{\bibfnamefont{M.~B.} \bibnamefont{Lundeberg}}
  \bibnamefont{and} \bibinfo{author}{\bibfnamefont{J.~A.} \bibnamefont{Folk}},
  \bibinfo{journal}{Nat. Phys.} \textbf{\bibinfo{volume}{5}},
  \bibinfo{pages}{894} (\bibinfo{year}{2009}).

\bibitem[{\citenamefont{Taylor and Heinonen}(2002)}]{taylor2002}
\bibinfo{author}{\bibfnamefont{P.~L.} \bibnamefont{Taylor}} \bibnamefont{and}
  \bibinfo{author}{\bibfnamefont{O.}~\bibnamefont{Heinonen}},
  \emph{\bibinfo{title}{A Quantum Approach to Condensed Matter Physics}}
  (\bibinfo{publisher}{Cambridge University Press}, \bibinfo{year}{2002}).

\bibitem[{\citenamefont{Beenakker and van Houten}(1991)}]{been1991}
\bibinfo{author}{\bibfnamefont{C.~W.~J.} \bibnamefont{Beenakker}}
  \bibnamefont{and} \bibinfo{author}{\bibfnamefont{H.}~\bibnamefont{van
  Houten}}, \bibinfo{journal}{Solid State Phys.} \textbf{\bibinfo{volume}{44}},
  \bibinfo{pages}{1} (\bibinfo{year}{1991}).

\bibitem[{\citenamefont{Tuinstra and Koenig}(1969)}]{tuinstra1969}
\bibinfo{author}{\bibfnamefont{F.}~\bibnamefont{Tuinstra}} \bibnamefont{and}
  \bibinfo{author}{\bibfnamefont{J.~L.} \bibnamefont{Koenig}},
  \bibinfo{journal}{J. Chem. Phys.} \textbf{\bibinfo{volume}{53}},
  \bibinfo{pages}{1126} (\bibinfo{year}{1969}).

\bibitem[{\citenamefont{Can\c{c}ado et~al.}()\citenamefont{Can\c{c}ado, Beams,
  and Novotny}}]{cancado2008B}
\bibinfo{author}{\bibfnamefont{L.~G.} \bibnamefont{Can\c{c}ado}},
  \bibinfo{author}{\bibfnamefont{R.}~\bibnamefont{Beams}}, \bibnamefont{and}
  \bibinfo{author}{\bibfnamefont{L.}~\bibnamefont{Novotny}},
  \bibinfo{journal}{Preprint at $<$http://arxiv.org/abs/0802.3709$>$}  (2008).

\bibitem[{\citenamefont{Casiraghi et~al.}(2009)\citenamefont{Casiraghi,
  Hartschuh, Qian, Piscanec, Georgi, Fasoli, Novoselov, Basko, and
  C.Ferrari}}]{casiraghi2009}
\bibinfo{author}{\bibfnamefont{C.}~\bibnamefont{Casiraghi {\it et al.}}},
  \bibinfo{journal}{Nano Lett.} \textbf{\bibinfo{volume}{9}},
  \bibinfo{pages}{1433} (\bibinfo{year}{2009}).

\bibitem[{\citenamefont{Ferrari and Robertson}(2001)}]{ferrari2001}
\bibinfo{author}{\bibfnamefont{A.~C.} \bibnamefont{Ferrari}} \bibnamefont{and}
  \bibinfo{author}{\bibfnamefont{J.}~\bibnamefont{Robertson}},
  \bibinfo{journal}{Phys. Rev. B} \textbf{\bibinfo{volume}{64}},
  \bibinfo{pages}{075414} (\bibinfo{year}{2001}).

\bibitem[{\citenamefont{Ferrari et~al.}(2006)\citenamefont{Ferrari, Meyer,
  Scardaci, Casiraghi, Lazzeri, Mauri, Piscanec, Jiang, Novoselov, Roth
  et~al.}}]{ferrari2006}
\bibinfo{author}{\bibfnamefont{A.~C.} \bibnamefont{Ferrari} {\it et al.}},
  \bibinfo{journal}{Phys. Rev. Lett.} \textbf{\bibinfo{volume}{97}},
  \bibinfo{pages}{187401} (\bibinfo{year}{2006}).

\bibitem[{\citenamefont{Can\c{c}ado et~al.}(2008)\citenamefont{Can\c{c}ado,
  Reina, Kong, and Dresselhaus}}]{cancado2008}
\bibinfo{author}{\bibfnamefont{L.~G.} \bibnamefont{Can\c{c}ado}},
  \bibinfo{author}{\bibfnamefont{A.}~\bibnamefont{Reina}},
  \bibinfo{author}{\bibfnamefont{J.}~\bibnamefont{Kong}}, \bibnamefont{and}
  \bibinfo{author}{\bibfnamefont{M.~S.} \bibnamefont{Dresselhaus}},
  \bibinfo{journal}{Phys. Rev. B} \textbf{\bibinfo{volume}{77}},
  \bibinfo{pages}{245408} (\bibinfo{year}{2008}).

\bibitem[{\citenamefont{Thomsen and Reich}(2000)}]{thomsen2000}
\bibinfo{author}{\bibfnamefont{C.}~\bibnamefont{Thomsen}} \bibnamefont{and}
  \bibinfo{author}{\bibfnamefont{S.}~\bibnamefont{Reich}},
  \bibinfo{journal}{Phys. Rev. Lett.} \textbf{\bibinfo{volume}{85}},
  \bibinfo{pages}{5214} (\bibinfo{year}{2000}).

\bibitem[{\citenamefont{Saito et~al.}(2001)\citenamefont{Saito, Jorio, Filho,
  Dresselhaus, Dresselhaus, and Pimenta}}]{saito2001}
\bibinfo{author}{\bibfnamefont{R.}~\bibnamefont{Saito} {\it et al.}},
\bibinfo{journal}{Phys. Rev. Lett.}
  \textbf{\bibinfo{volume}{88}}, \bibinfo{pages}{027401}
  (\bibinfo{year}{2001}).

\bibitem[{\citenamefont{Can\c{c}ado et~al.}(2004)\citenamefont{Can\c{c}ado,
  Pimenta, Neves, Dantas, and Jorio}}]{cancado2004}
\bibinfo{author}{\bibfnamefont{L.~G.} \bibnamefont{Can\c{c}ado} {\it et al.}},
  \bibinfo{journal}{Phys. Rev. Lett.} \textbf{\bibinfo{volume}{93}},
  \bibinfo{pages}{247401} (\bibinfo{year}{2004}).

\bibitem[{\citenamefont{Lucchese et~al.}(2010)\citenamefont{Lucchese, Stavale,
  Ferreira, Vilani, Moutinho, Capaz, Achete, and Jorio}}]{lucchese2010}
\bibinfo{author}{\bibfnamefont{M.~M.} \bibnamefont{Lucchese} {\it et al.}},
  \bibinfo{journal}{Carbon} \textbf{\bibinfo{volume}{48}},
  \bibinfo{pages}{1592} (\bibinfo{year}{2010}).

\bibitem[{\citenamefont{Basko}(2009)}]{basko2009}
\bibinfo{author}{\bibfnamefont{D.~M.} \bibnamefont{Basko}},
  \bibinfo{journal}{Phys. Rev. B} \textbf{\bibinfo{volume}{79}},
  \bibinfo{pages}{205428} (\bibinfo{year}{2009}).

\bibitem[{\citenamefont{Basko et~al.}(2009)\citenamefont{Basko, Piscanec, and
  Ferrari}}]{baskoprb}
\bibinfo{author}{\bibfnamefont{D.~M.} \bibnamefont{Basko}},
  \bibinfo{author}{\bibfnamefont{S.}~\bibnamefont{Piscanec}}, \bibnamefont{and}
  \bibinfo{author}{\bibfnamefont{A.~C.} \bibnamefont{Ferrari}},
  \bibinfo{journal}{Phys. Rev. B} \textbf{\bibinfo{volume}{80}},
  \bibinfo{pages}{165413} (\bibinfo{year}{2009}).

\end{thebibliography}

\end{document}